\documentclass[aps,twocolumn,showpacs,superscriptaddress,amsmath,amssymb,showkeys,longbibliography]{revtex4-1}

\usepackage{graphicx}
\usepackage{braket}
\usepackage{hyperref}
\usepackage{epstopdf}

\begin{document}

\title{Alleviating Oscillatory Approximate-Kernel Solutions for Cylindrical Antennas Embedded in a Conducting Medium: a Numerical and Asymptotic Study}

\author{Th. K. Mavrogordatos}
\email[Email address: ]{themis.mavrogordatos@fysik.su.se}
\affiliation{Department of Physics, Stockholm University, SE-106 91, Stockholm, Sweden}
\author{C. Mystilidis}
\affiliation{School of Electrical and Computer Engineering, National Technical University of Athens, Athens, 15 780 Zografou, Greece}
\author{P. J. Papakanellos}
\affiliation{Department of Aeronautical Sciences, Hellenic Air Force Academy, Dekelia Air Force Base, Athens 13 671, Greece}
\author{G. Fikioris}
\affiliation{School of Electrical and Computer Engineering, National Technical University of Athens, Athens, 15 780 Zografou, Greece}

\date{\today}

\begin{abstract}
We alleviate the unnatural oscillations occurring in the current distribution along a linear cylindrical antenna center-driven by a delta-function generator and embedded in a conducting medium. The intensely fluctuating current arises as a small-$z_0$ asymptotic (or numerical) solution of the classical integral equations of antenna theory, for a cylindrical dipole of infinite (or finite) length, where $z_0$ is the discretization length. To alleviate the oscillations, we employ an appropriate effective current further to the recent remedy of oscillations attained for a perfectly conducting linear cylindrical antenna of finite length for the case where the surrounding medium is free space. We derive asymptotic formulas for the infinite antenna, which are then put to numerical test. Furthermore, we point to the physical significance of the effective current whose function transcends a mere computational device.
\end{abstract}

\pacs{84.40.Ba, 03.50.De, 41.20.Jb, 02.70.-c}
\keywords{Integral equation methods, Hall\'{e}n's integral equation, approximate kernel, method of moments, effective current, antennas in matter}

\maketitle

\section{Introduction}

Somewhat more and somewhat less than a century has lapsed since Pocklington \cite{Pocklington1898} and Hall\'{e}n \cite{Hallen1938}, respectively, published reports on the two classical integral equations that currently bear their names (another relevant, perhaps less-known, early work is \cite{LVKing}). These integral equations are satisfied by the current distribution on a straight, thin, center-driven wire. Today, there are two main possibilities for the kernel featuring in these equations, the exact and the approximate (or reduced) kernel, with serious consequences on the behavior of the respective solutions. Many early developments are addressed in King's comprehensive treatise of 1956 \cite{King1956}, shortly before the advent of numerical methods in the mid-1960s \cite{harringtonbook, CollinZucker}. The application of moment methods to thin-wire integral equations alongside their associated difficulties has a long history (see e.g., \cite{Strait1980} and references therein, as well as \cite{CollinEquiv1984}). In 2001, these difficulties were described in detail and attributed to the nonsolvability of the integral equation itself, rather than to the deficiency of the numerical method employed \cite{Hallen2001, noslution2001GF}; these discussions have since been corroborated for a variety of antenna geometries and driving configurations. At present, Hall\'{e}n's and Pocklington's equations remain a subject of active research. A representative number of recent works concerning the exact kernel are given in \cite{EffCurrent2011}, with a very efficient solution standing out in \cite{BrunoSIAM}. At the same time, the approximate, non-singular kernel enjoys wide popularity in the literature \cite{balanis1989, stutzman1998, kraus2002, balanis2005}, as well as in standard antenna analysis software \cite{NEC2001}. 

In this paper, we focus on a method of post-processing unphysical numerical solutions obtained with the approximate kernel when the discretization length is small. The first application of the effective current on a current distribution obtained via a Galerkin formulation was discussed in \cite{papakanellos2007possible} in the spirit of the Method of Auxiliary Sources (MAS) for closed scattering surfaces \cite{PapakanellosCapsalis2004}, an assumption not strictly applicable to our antenna in question. The authors remark that ``although oscillations in the expansion coefficients and the associated line currents are unavoidable in approximate-kernel formulations of thin-wire antennas, meaningful and  useful current distributions can be derived by calculating a smoothed current through the magnetic  field at a radial distance away from the line current.'' This idea, central in our analysis, revolves around calculating the magnetic field generated by a line current \textemdash{this} current being the solution of an antenna equation with the approximate kernel \textemdash{at} a distance equal to the antenna radius. The field is then associated with a surface current density and ultimately with the effective current as the density times the circumference. 

Appealing to auxiliary formulations in order to circumvent the nonsolvability of a governing equation finds an application in various branches of physics. For example, in \cite{Cade1994}, Cade proposes an alternative integral equation to replace a commonly used equation in electrostatics, and by extension to the slender-body theory of hydrodynamics (see also the asymptotic analysis of \cite{handelsman_keller_1967}), with no solution. He does so by means of a ``numerical or asymptotic analysis'' to approximate the solution of the problem, which the original equation fails to yield ``through being meaningless'' (see also \cite{FikiorisElectrostatics2017}). Today, theoretical investigations of electromagnetic and light-scattering problems are commonplace. In the widely used MAS and its numerous implementations, one typically seeks electric current sources (which are intermediate results arising from a solution of a linear algebraic system) located on a closed auxiliary surface within an impenetrable scatterer, such that the proper boundary conditions of the problem are satisfied. The field produced by these sources (the so-called MAS field) is an approximation of the true scattered field. Although the MAS currents exhibit rapid and unphysical oscillations, the field they produce is smooth and free of serious difficulties. Furthermore, the simplicity of common scattering problems dealt with MAS allow one to demonstrate analytically that the large$-N$ limit of the MAS field (where $N$ is the number of auxillary current-sources) yields the true scattered field. The similarity between the oscillatory behavior of the auxiliary currents in MAS and the currents in radiators of superdirective arrays was noted in the pioneering work of \cite{Malakshinov1977}, while the authors of \citep{Andrianesis2012} derive asymptotic formulas for the near field that allow one to distinguish the superdirective-type behaviors in MAS from effects due to roundoff. At the same time, the properties of rapidly-oscillating currents and associated rapidly-decaying near fields akin to superdirectivity are investigated in \cite{EffCurrent2013}; this work is dedicated to the asymptotic analysis of the effective current, as introduced in \cite{EffCurrent2011} for the perfectly-conducting antenna of infinite length. For a detailed discussion on convergent fields generated by divergent currents in the MAS, we refer the reader to \cite{GMTFikiorisTsitsas}, which is specifically devoted to light scattering.

Antennas radiate in matter (see Ch. 7 of \cite{antennasinmatter}); the physical significance of the effective current ought then to be assessed against a background of finite conductivity. Our investigation is motivated by a recent work \cite{Mistilidis2019} reporting on the modification of the current obtained by solving Hall\'{e}n's equation when the surrounding medium becomes imperfectly conducting. In Sec. \ref{sec:EquationsKernels}, we will define Pocklington and Hall\'{e}n equations, proceeding to a formulation of the effective current with recourse to the infinite antenna in matter in Sec. \ref{sec:effectiveinf}. After a short discussion revealing the physical meaning of such a post-processing method based on Maxwell's equations and the usual boundary conditions for the electromagnetic field in Sec. \ref{sec:physicalsign}, we will present numerical results for the finite-length antenna in Sec. \ref{sec:numerics}, compared against our preceding asymptotic analysis for the infinite antenna. We apply two versions of the method of moments, assessing their numerical stability, based on which we compute effective-current distributions. Throughout our analysis, we aim to demonstrate the independence of the discussed phenomena on the particulars of a numerical technique while comparing the output with available experimental results in the literature.

\section{Pocklington and Hall\'{e}n equations for a linear antenna in a conducting medium}
\label{sec:EquationsKernels}

\subsection{Fredholm Integral Equation of the First Kind}

In this section, we will define the well-known classical electromagnetic equations yielding the current distribution on a linear cylindrical perfectly conducting (PC) antenna driven by a delta-function generator (DFG) [an infinitesimal gap at the center across which a scalar potential difference Re$\{Ve^{-i\omega t}\}$ is maintained]. Our notation and conventions follow closely those of \cite{EffCurrent2011, EffCurrent2013, anastasiscnt}. For an {\it electrically and geometrically thin} cylindrical antenna of radius $a$ and finite length $2h$ such that $ka \ll 1$ and $a/h \ll 1$ (where $k$ is the free-space wavenumber), there exists a current density on both the inside and the outside tubular surfaces. The total current $I(z)$ is defined as the sum of these two current densities multiplied by $2\pi a$, the circumference of the cylindrical tube. $I(z)$ satisfies Pocklington's equation (PE),  
\begin{equation}\label{eq:Pocklington}
\left(\frac{\partial^2}{\partial z^2} + k_c^2\right)\int_{-h}^{h} K(a, z-z^{\prime})I(z^{\prime})dz^{\prime}=\frac{iV k_c}{\zeta_c}\delta(z),
\end{equation}
or, equivalently, its Fredholm-type form known as Hall\'{e}n's equation (HE),
\begin{equation}\label{eq:Hallen}
\int_{-h}^{h} K(a, z-z^{\prime})I(z^{\prime})dz^{\prime}=\frac{iV}{2\zeta_c}\sin(k_c|z|)+C \cos(k_cz),
\end{equation}
respectively, where $-h<z<h$. In the above equations, $k_c \equiv \omega \sqrt{\mu \epsilon_c}$ is the complex wavenumber, $\zeta_c \equiv \sqrt{\mu/\epsilon_c}$ is the complex impedance of the surrounding medium, with dielectric permittivity $\epsilon_c \equiv \epsilon (1+i \tan \delta)$ defining the {\it loss tangent} $\tan \delta \equiv \gamma/(\omega \epsilon)$. The magnetic permeability is always taken here equal to its free-space value $\mu=\mu_0$, and \textemdash{unless} explicitly stated otherwise \textemdash{we} take $\epsilon=\epsilon_0$. With $k=2\pi/\lambda=\omega/c$ and $\zeta_0=\sqrt{\mu_0/\epsilon_0}=120\pi\,$Ohms the free-space wavenumber and intrinsic impedance, respectively, we note that $(k_c/\zeta_c)/(k/\zeta_0)=1+i\tan \delta$. In Eq. \ref{eq:Hallen}, the complex constant $C$ is determined by means of the approximate boundary condition for the total current at the two ends of the antenna, where \textemdash{as} is standard practice \textemdash{we} assume that the current flowing on the outer surface turns inwards to the inner surface,
\begin{equation}\label{eq:edgecondition}
I(\pm h)=0.
\end{equation}
The condition of Eq. \ref{eq:edgecondition} is best viewed as the condition at the edge of a tubular antenna: the current that flows on the outer surface turns in toward the inner surface. When the surrounding medium has a finite conductivity, the entire formulation is put to question since a current leaks out towards the surrounding medium, both at the ends but also along the entire length of the antenna. In fact, the leakage along the length is clearly the decisive effect; as such, it is explicitly discussed in \cite{KingHarrison} where the authors, after having discussed leakage, go on to formulate and solve precisely the integral equation \ref{eq:Hallen} with the edge condition \ref{eq:edgecondition}. The authors of \cite{PopovicBook} also state firmly that ``...the presence of losses in a (homogeneous) medium leaves all the equations formally intact, but $k$ and $\zeta_0$ (in the present notation) become complex.'' An asymptotic solution for the current near the ends of a finite-length antenna that, via the Wiener-Hopf method, shows a divergence from the usual sinusoidal current distribution, is presented in \cite{ShenWu}.

\subsection{Exact and Approximate Kernels}

In the bibliography, we encounter two very popular choices for the kernel $K(z)$ in Eqs. \ref{eq:Pocklington} and \ref{eq:Hallen}, namely the {\it exact} and the {\it approximate} kernels \footnote{In our work, we denote by the subscripts (ex, ap) the quantities corresponding to the (exact, approximate) kernels.}. The approximate kernel replaces all distances between the source and field points appearing in the integral for the exact kernel by the radius $a$, hence placing the current on the $z$-axis, which is the axis of symmetry. In practice, this means that a surface current at $\rho=a$ is approximated by a line current at $\rho=0$, pointing to the nonsolvability of HE with the approximate kernel \cite{CollinZucker, Hallen2001}. For the PC antenna in a conducting medium, the exact kernel
\begin{equation}\label{eq:Exkernels}
K_{\rm ex}(a,z)=\frac{1}{8\pi^2} \int\limits_{-\pi}^{\pi} \frac{e^{i k_c R(z,\phi;a)}}{R(z,\phi;a)}d\phi,
\end{equation}
with $R(z,\phi;a)\equiv\sqrt{z^2+4a ^2\sin^2\left(\frac{\phi}{2}\right)}$, is thus frequently replaced by the approximate kernel,
\begin{equation}\label{eq:ApprKernel}
K_{\rm ap}(a,z)=\frac{1}{4\pi} \frac{e^{i k_c \sqrt{z^2+a^2}}}{\sqrt{z^2+a^2}},
\end{equation}
a simplification which can be regarded as a non-singular version of the exact kernel. In fact, $K_{\rm ap}(a,z)$ is an analytic function of the variable $z$. 

\begin{figure*}
\begin{center}
\includegraphics[width=\textwidth]{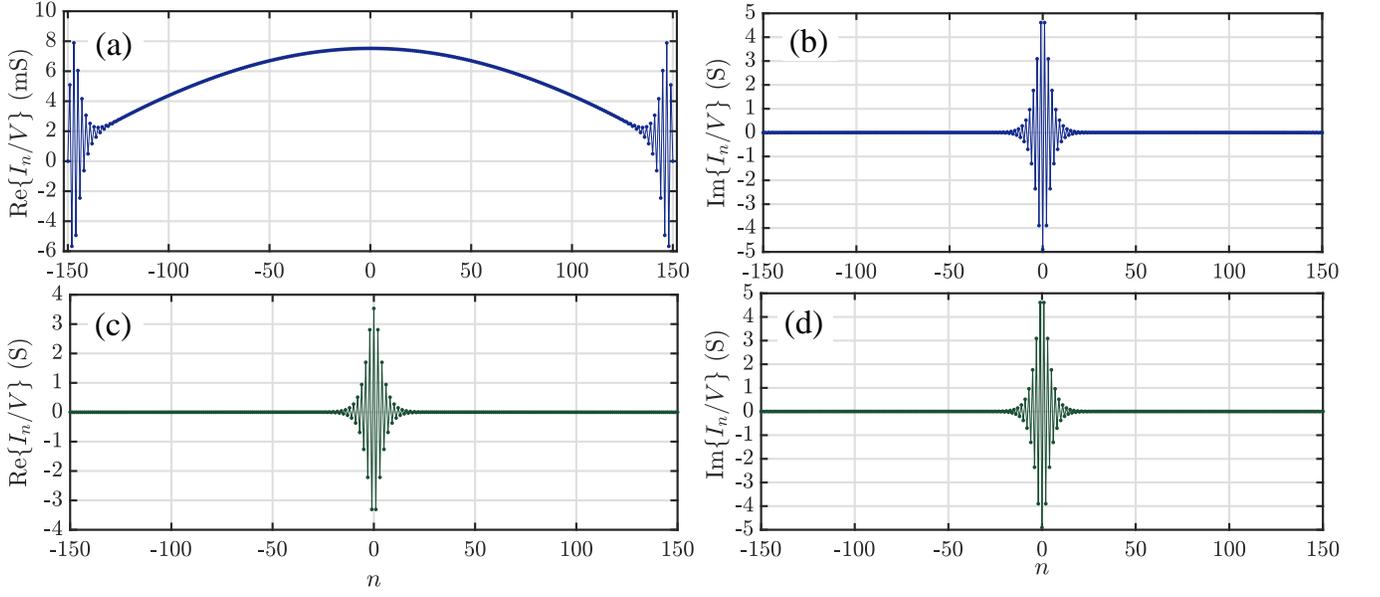}
\end{center}
\caption{{\it Oscillatory current distributions $I_{n, {\rm ap}}/V$ with the approximate kernel on a half-wavelength dipole.} Real {\bf (a)} and imaginary {\bf (b)} part of the current distribution obtained from the numerical solution of Eq. \ref{eq:Hallen} with the approximate kernel in free space ($\tan \delta=0$). Real {\bf (c)} and imaginary {\bf (d)} part of the current distribution obtained from the numerical solution of Eq. \ref{eq:Hallen} with the approximate kernel in a conducting medium with $\tan \delta=0.72$ \footnote{In most dielectrics at microwave frequencies, one typically encounters $\tan \delta \ll 1$ (see e.g., \cite{dielectrictand, ceramics}) while studies in conducting media report $\tan \delta \sim 1$ in the MHz range \cite{PopovicConducting, IizukaKing}.}. The remaining parameters are: $h/\lambda=0.25$, $a/\lambda=0.007022$, $N=150$. MoM-A has been used here to generate the depicted results.}
\label{fig:approx150}
\end{figure*}

In the free-space case, using the approximate kernel leads to the appearance of unphysical oscillations in the real part of the current distribution at the ends of a finite-length center-driven antenna, and at both the center and the ends of the antenna in the imaginary part, as discussed in detail in \cite{Hallen2001}. These oscillations appear for $z_0 \approx h/N \ll a$, where $N$ is the number of discretization points and $z_0$ is the discretization length. We remark  that the central oscillations of the imaginary part are much more acute than the oscillations at the antenna ends (note the scale difference in Figs. \ref{fig:approx150}(a) and \ref{fig:approx150}(b)). For an infinite antenna, one sends both $h$ and $N$ simultaneously to infinity keeping $z_0$ fixed, while the position along the $z$-axis \textemdash{corresponding} to the expansion coefficient $I_n^{(\infty)}$ of the discretized current \textemdash{is} $nz_0$ ($n=0, \pm 1, \pm 2, \ldots$). In this work, we further assume that $|k_c z_0| \ll 1$ and $|k_c a| \ll 1$. Milder and less rapid oscillations occur when one employs an equally non-singular kernel, the so-called {\it extended} kernel, produced by applying a differential operator to the approximate kernel \cite{extended2016}. A remedy of these oscillations is addressed in detail in \cite{EffCurrent2011, EffCurrent2013}. Further along that path of investigation, a recent publication \cite{Mistilidis2019} concluded that, for an infinite antenna in which $z_0 \to 0$ and $nz_0/a = \mathcal{O}(1)$,
\begin{equation}\label{eq:infinitetanddelta}
\frac{I_{n,{\rm ap;\, conducting}}^{(\infty)}}{V} \sim (1+i \tan \delta) \frac{I_{n,{\rm ap;\, free\,\, space}}^{(\infty)}}{V},
\end{equation}
where $I_{n,{\rm ap;\, free space}}^{(\infty)}/V$ is a purely imaginary current distribution pertaining to the infinite-length version of Eq \ref{eq:Hallen}. For the case of a {\it finite} antenna, Eq. \ref{eq:infinitetanddelta} suggests the additional appearance of central oscillations in the real part with the oscillations in the imaginary part remaining roughly the same (for a detailed comparison see Table I of \cite{Mistilidis2019}). In the next section, we will therefore aim to extend the definition of the effective current to tackle the onset of oscillatory kernel solutions in a conducting medium when the approximate kernel is used. 

\section{The effective current for an infinite antenna}
\label{sec:effectiveinf}

In this section, we discuss the effective current for the infinite antenna and derive certain asymptotic formulas, which will prove useful in the numerical analysis. We denote by $I_{n}$, $n=0, \pm1, \pm 2, \ldots$, the basis functions coefficients, with $I_{0}$ located at $z=0$. Setting the length $h$ in the limits of the integral in Eq. \ref{eq:Pocklington} to infinity ($h=\infty$), we look for a current distribution of the form 
\begin{equation}\label{eq:CurrentSinus}
I(z) \simeq \sum_{n=-\infty}^{\infty} I_{n}^{(\infty)}s(z-nz_0),
\end{equation}
where the sinusoidal basis function $s(z)$ of length $2z_0$ has the form
\begin{equation}\label{eq:sinbasis}
s(z)=\begin{cases}\displaystyle\frac{\sin[k_c (z_0-|z|)]}{\sin(k_c z_0)},\quad &-z_0\leq z \leq z_0, \\ 0, \quad &|z|> z_0. \end{cases}
\end{equation}
We substitute Eq. \ref{eq:sinbasis} into Eq. \ref{eq:Pocklington} (with $h=\infty$) and take the inner product with (displaced) triangular functions $t(z-lz_0)$ of length $2z_0$ [such that $t(0)=1$ and $t(\pm z_0)=0$], defined as \cite{Hallen2001, EffCurrent2011}
\begin{equation}\label{eq:triangular}
t(z)=\begin{cases}\displaystyle\frac{z_0-|z|}{z_0},\quad &-z_0\leq z \leq z_0, \\ 0, \quad &|z|> z_0. \end{cases}
\end{equation}
This yields an infinite system of equations for the coefficients $I^{(\infty)}_{n}$. These coefficients are used in turn to define the \textit{effective current} for the range $0<\rho \leq a$ (and not only for $\rho=a$), via the magnetic field of the sinusoidal current $I_{n}^{(\infty)} s(z-nz_0)$, as (see \cite{EffCurrent2011, EffCurrent2013})
\begin{equation}\label{eq:IeffIn}
\begin{aligned}
I_{\rm eff}^{(\infty)}(\rho, z, z_0)&=\frac{1}{2i \sin(k_c z_0)}\\
& \times \sum_{n=-\infty}^{\infty}[f_{n+1}-2\cos(k_c z_0)f_{n}+f_{n-1}]I_{n}^{(\infty)},
\end{aligned}
\end{equation}
with $f_{n}\equiv\exp[ik_c\sqrt{(nz_0-z)^2+\rho^2}]$. Intuitively, we can think of the coefficients $I_{n}^{(\infty)}$ as a line current on the $z$-axis creating a magnetic field $H_{\phi}(\rho,z)$ hence counteracting the usage of the approximate kernel; we will return to this point in Sec. \ref{sec:physicalsign}. Furthermore, asymptotic expressions for $I_{\rm eff}(\rho=a, z ,z_0)$ in the limit of small discretization length $z_0 \to 0$ have been obtained in \cite{EffCurrent2011} to provide a link with the logarithmically diverging exact-kernel current distribution for $\rho=a$ as $z\to 0$, a property that will concern us in Sec. \ref{sec:numerics}. 

We will now follow a series of manipulations in parallel to Sec. 3 of \cite{EffCurrent2011}. The effective current for the infinite antenna with a nonzero $z_0\ll a$ is found to be given by the expression
\begin{equation}\label{eq:IeffF}
I_{\rm eff}(\rho, z, z_0)=\frac{1}{2\pi}\int_{0}^{\pi} \frac{F(\theta, \rho, z, z_0) + F(-\theta, \rho, z, z_0)}{A(\theta, z_0)}\, d\theta,
\end{equation}
where 
\begin{equation}
F(\theta, \rho, z, z_0) \equiv\frac{V z_0}{2\zeta_c}\sum_{l=-\infty}^{\infty}e^{i k_c \sqrt{(lz_0-z)^2+\rho^2}}\,e^{-i l \theta}.
\end{equation}
Using the Poisson summation formula, this quantity can be alternatively written as [in this paper, $\overline{F}(\zeta) \equiv \int_{-\infty}^{\infty}F(z)e^{i\zeta z}\, dz$ defines the Fourier transform of the function $F(z)$]
\begin{equation}\label{eq:Ftheta}
F(\theta, \rho, z, z_0)= \frac{V}{2\zeta_c}\sum_{m=-\infty}^{\infty}e^{i\frac{2\pi m-\theta}{z_0}z}\overline{L}\left(\rho, \frac{2\pi m -\theta}{z_0} \right),
\end{equation}
with
\begin{equation}\label{eq:LFT}
\overline{L}(\rho, \zeta)=\frac{2k_c\rho}{i\sqrt{\zeta^2-k_c^2}}K_{1}\left(\rho \sqrt{\zeta^2-k_c^2}\right)
\end{equation}
and $K_1$ the modified Bessel function of the second kind and first order.

From \cite{Mistilidis2019} and \cite{Hallen2001}, the denominator in Eq. \ref{eq:IeffF} assumes the form
\begin{equation}\label{eq:Athetaz0}
A(\theta, z_0)=z_0 \sum_{m=-\infty}^{\infty} \overline{K}_{\rm ap}\left(a, \frac{2 m\pi-\theta}{z_0} \right) \frac{\sin^2(\frac{\theta}{2})}{(m\pi-\frac{\theta}{2})^2},
\end{equation} 
where
\begin{equation}\label{eq:KFTr}
\overline{K}_{\rm ap}(a,\zeta)=\frac{1}{2\pi}K_0\left(a \sqrt{\zeta^2-k_c^2}\right),
\end{equation}
with $K_0$ the modified Bessel function of the second kind and zero order. Substituting now Eqs. \ref{eq:Athetaz0}, \ref{eq:LFT} and \ref{eq:Ftheta} in Eq. \ref{eq:IeffF}, and using the change of variable $\theta=\pi-\phi$, yields
\begin{equation}\label{eq:Ieffphi}
\begin{aligned}
&I_{\rm eff}(\rho, z, z_0) \simeq \frac{(-1)^n}{4\pi z_0}\\
&\times \int_{0}^{\pi} \frac{G(\pi-\phi, \rho, z_0)}{\cos^2(\frac{\phi}{2})\left[D_{+}(\phi,z_0) + D_{-}(\phi,z_0) \right]}\cos(n\phi)\, d\phi,
\end{aligned}
\end{equation}
where $D_{\pm}(\phi,z_0)\equiv (\pi\pm\phi)^{-2}\overline{K}_{\rm ap}\left(a, \frac{\pi\pm\phi}{z_0}\right)$. We note that in Eq. \ref{eq:Ieffphi} the integration contour is {\it not} indented at $\phi=\pi-k_c z_0$ since $k_c$ is complex. In the integrand, setting $x=\phi a/z_0$, we write [see also Eqs. (1.8), (2.1), (2.9), (2.10) of \cite{EffCurrent2013}]
\begin{equation}\label{eq:Bexpr}
\begin{aligned}
&G\left(\pi - \frac{z_0 x}{a}, \rho, z_0\right)=G\left(\frac{z_0 x}{a}-\pi, \rho, z_0\right)\\
&=\frac{V}{2\zeta_c}\sum_{m=-\infty}^{\infty}\overline{L}\left(\rho, \frac{(2m+1)\pi - \frac{z_0 x}{a}}{z_0}\right)\\
&=-i \frac{V}{\zeta_c}k_c z_0^2\frac{\rho}{z_0}\sum_{m=-\infty}^{\infty}\frac{1}{r_{m}}K_{1}\left(\frac{\rho}{z_0}r_{m}\right),
\end{aligned}
\end{equation}
with 
\begin{equation}
\begin{aligned}
r_{m}&\equiv \sqrt{\left[(2m-1)\pi-\frac{z_0 x}{a}\right]^2-(k_c z_0)^2}\\
& \simeq \left|(2m-1)\pi -\frac{z_0 x}{a}\right|,
\end{aligned}
\end{equation}
where, in the last step, we have omitted the term $(k_c z_0)^2$ by virtue of $|k_c z_0| \ll 1$. 

From this point onwards, our derivation proceeds with identical steps to those taken in \cite{EffCurrent2013}, since the ratio $k_c/\zeta_c$ carries over and replaces $k/\zeta_0$ in all expressions resulting from Eq. \ref{eq:Bexpr}. In particular, for $a/z_0 \to \infty$, using the asymptotic formula $\overline{K}_{\rm ap}(a,\zeta) \sim (1/2)\sqrt{1/(2\pi a |\zeta|)}\,e^{-a|\zeta|}$ as $|\zeta| \to \infty$, we make the approximation $D_{\pm} \sim e^{\pm x} (\pi \mp z_0 x/a)^{5/2}$ with $(D_{+}+D_{-})\sim 2\pi^{5/2} \cosh{x}$. Hence, from Eq. \ref{eq:Ieffphi} we are left with the integral $\int_{0}^{\infty} [\cos(n z_0 x/a)/\cosh{x}]dx$, tabulated as Entry 2.5.46.5 of \cite{Prudnikov}.

From the above considerations, an asymptotic formula predicting rapid oscillations can be readily established for $\rho/z_0 \sim 1$ and $nz_0/a=\mathcal{O}(1)$, i.e., near the driving point in the axial direction and  infinitesimally close to the axis of the tubular antenna in the radial direction, 
\begin{equation}\label{eq:asymptoticIeff}
\begin{aligned}
I_{{\rm eff}} (\rho ,nz_{0} ,z_{0})&\sim -i(-1)^{n}\frac{V}{\zeta _{c}} \frac{\pi ^{3}}{16\sqrt{2}} k_c z_{0} \sqrt{\frac{z_{0}}{a}}  \\
&\times \exp \left(\pi \frac{a}{z_{0}}\right)\frac{1}{\cosh \left(\displaystyle\frac{\pi nz_{0}}{2a}\right)}\eta\left(\frac{\rho}{z_{0}}\right),
\end{aligned}
\end{equation}
where
\begin{equation*}
\eta(y) \equiv \frac{8y}{\pi}\sum_{m=1}^{\infty}\frac{1}{2m-1}K_{1}[(2m-1)\pi y], \quad y\geq 0,
\end{equation*}
with $\eta(0)=1$. The effective-current distribution is oscillatory due to the presence of the factor $(-1)^n$, while oscillations are intense due to the presence of the term $\exp(\pi a /z_0)$. The resulting expression differs only by a multiplicative factor of $(k_c/\zeta_c)/(k/\zeta_0)$ in comparison to the lossless case. This abides with the prediction of Eq. \ref{eq:infinitetanddelta}. When moving further away from the $z$-axis, with $z_0 \ll \rho \ll a$ and $nz_0/a=\mathcal{O}(1)$, oscillations in the effective current persist but with decreasing magnitude, following \cite{EffCurrent2013}
\begin{equation}\label{eq:Ieffasymlargerrho}
\begin{aligned}
I_{\rm eff}(\rho, nz_0, z_0)& \sim -i (-1)^{n} \frac{\pi^2 V}{2\zeta_c}k_c z_0 \sqrt{\frac{\rho}{a}}\exp\left[\frac{\pi(a-\rho)}{z_0}\right] \\
&\times  \cos\left(\frac{\pi \rho}{2a}\right) \displaystyle\frac{\cosh\left(\displaystyle\frac{n\pi z_0}{2a}\right)}{\cosh\left(\displaystyle\frac{n\pi z_0}{a}\right) + \cos\left(\displaystyle\frac{\pi \rho}{a}\right)},
\end{aligned}
\end{equation}
to produce eventually a non-oscillatory current distribution on the surface $\rho=a$ \citep{EffCurrent2013}. The alleviation of oscillations is verified in our numerical treatment of Sec. \ref{sec:numerics} for a dipole of finite length.
\begin{figure}
\begin{center}
\includegraphics[width=0.5\textwidth]{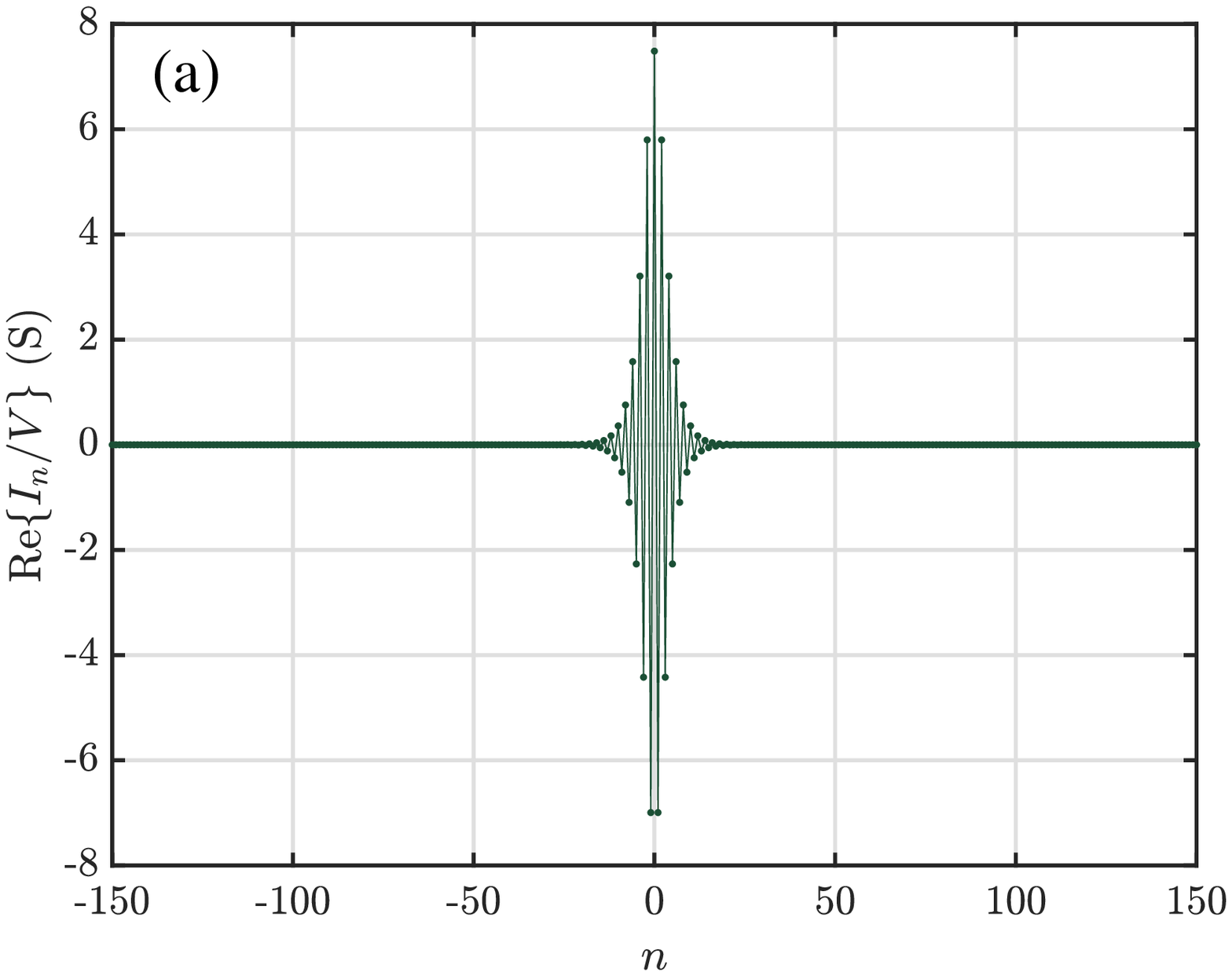}
\includegraphics[width=0.5\textwidth]{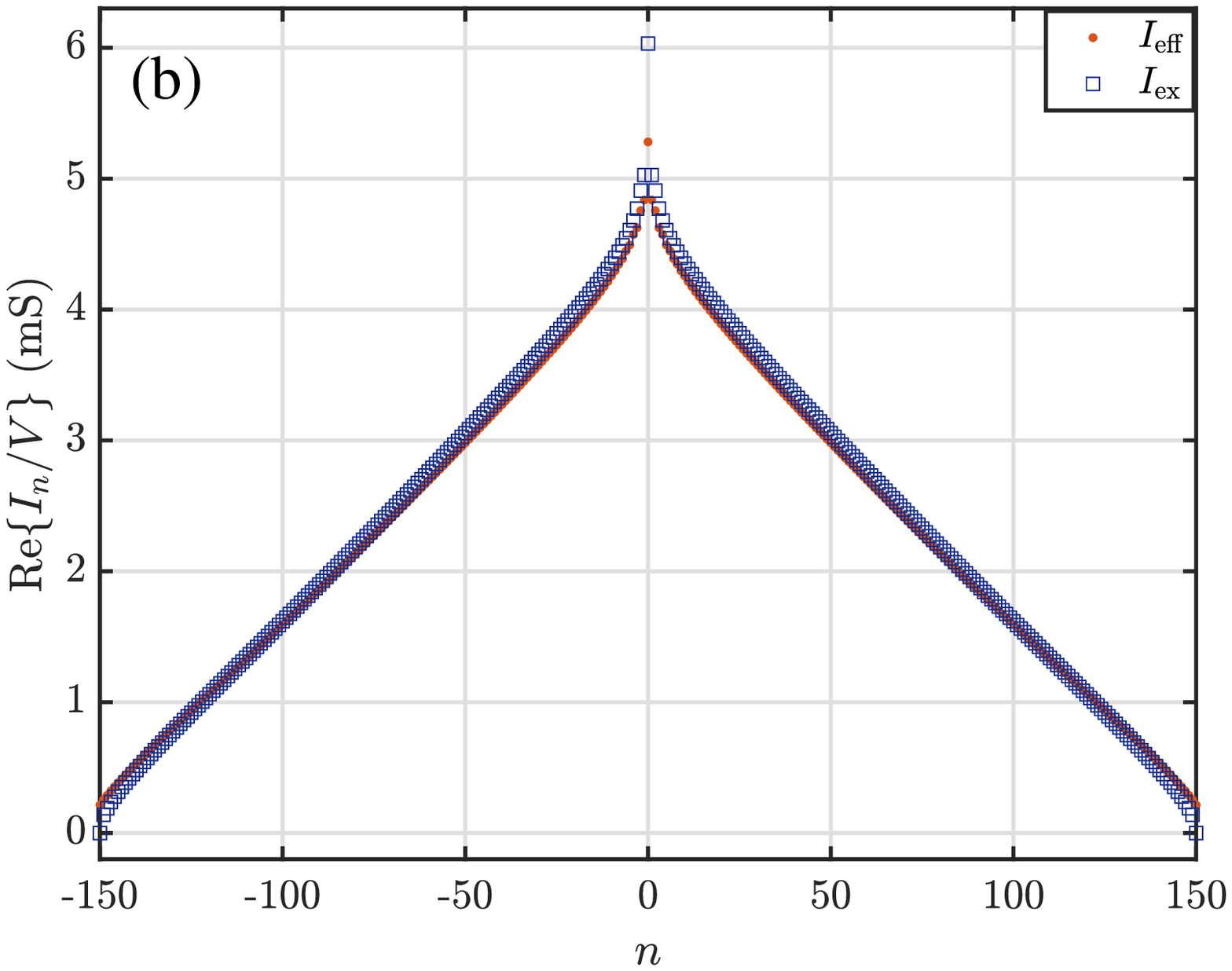}
\end{center}
\caption{{\it Effective current as a function of the radial distance.} {\bf (a)} Asymptotic current distribution of the effective current {\it on the antenna axis}, using Eq. \ref{eq:asymptoticIeff} with $\rho \sim z_0 \to 0$. The remaining parameters are: $h/\lambda=0.25$, $a/\lambda=0.007022$, $N=150$, $\tan \delta=0.72$.  {\bf (b)} Real part of the effective current distribution $I_{n, {\rm eff}}/V$ (in orange dots), calculated from the Hertzian dipole approximation (see discussion below Eq. \ref{eq:IeffInFin}) post-processing the oscillating current distribution $I_{n, {\rm ap}}/V$ obtained from the numerical solution of Eq. \ref{eq:Hallen} with the approximate kernel, superimposed on top of the numerical solution $I_{n, {\rm ex}}/V$ of Eq. \ref{eq:Hallen} with the exact kernel (in blue squares) {\it at $\rho=a$}. We used MoM-A to generate the results depicted in frame (a).}
\label{fig:effectivetwodist}
\end{figure}

\section{On the physical significance of the effective current}
\label{sec:physicalsign}

Further to our discussion on the effective current in Sec. \ref{sec:effectiveinf}, we will now solve a standard boundary value problem for the infinite antenna with the aim of determining the surface-current density at $\rho=a$. Our treatment concerns both the interior and the exterior of the tube, denoted by the superscripts $\text{(in)}$ and $\text{(out)}$, respectively. First, we consider Helmholtz equation for the electric-field component $E_z(\rho,z)$, parallel to the axis of the linear cylindrical dipole in question, 
\begin{equation}\label{eq:HelmholtzEz}
(\nabla^2 + k_c^2)E_{z}(\rho, z)=0.
\end{equation} 
Taking the Fourier transform with respect to $z$ we obtain
\begin{equation}\label{eq:FourierHelmholtz}
\left[\frac{1}{\rho}\frac{\partial}{\partial \rho}\left(\rho \frac{\partial}{\partial \rho}\right)-\zeta^2 + k_c^2\right]\overline{E}_{z}(\rho, \zeta)=0
\end{equation}
yielding
\begin{equation}\label{eq:BesselFT}
\rho^2 \frac{\partial^2}{\partial \rho^2}\overline{E}_{z}(\rho, \zeta) + \rho \frac{\partial}{\partial \rho}\overline{E}_{z}(\rho, \zeta) + \rho^2 (k_c^2-\zeta^2)\overline{E}_{z}(\rho, \zeta)=0.
\end{equation}
This is an equation of Bessel type. The boundary condition imposed by the DFG reads
\begin{equation}\label{eq:BCDFG1}
E_{z}^{\rm (in)}(a,z)=E_{z}^{\rm (out)}(a,z)=-V \delta(z),
\end{equation}
which transforms to
\begin{equation}\label{eq:BCDFG2}
\overline{E}_{z}^{\rm (in)}(a, \zeta)=\overline{E}_{z}^{\rm (out)}(a, \zeta)=-V.
\end{equation}
For $\rho <a$, the physically acceptable solution of Eq. \ref{eq:BesselFT} (which remains finite on the $z$-axis) subject to the boundary condition of Eq. \ref{eq:BCDFG2} reads \footnote{For the case of free space, we note that the requirement of standing waves on the inner metal surface of the tubular antenna is fulfilled when $ka<\rho_1^0$, where $\rho_1^0$ is the first zero of the Bessel function of the first kind and zero order, $J_{0}(\rho_1^0)=0$. This condition is {\it a priori} met since $ka \ll 1$.}
\begin{equation}\label{eq:SolInside}
\overline{E}_{z}^{\rm (in)}(\rho, \zeta)=-V \frac{I_{0}(\rho\sqrt{\zeta^2-k_c^2})}{I_{0}(a\sqrt{\zeta^2-k_c^2})},
\end{equation}
where $I_0$ is the modified Bessel function of the first kind and zero order.

For $\rho>a$, we require a solution of the Bessel differential equation that satisfies the Sommerfeld radiation condition, retaining only the Hankel function of the first kind, with $H_0^{(1)}(x) \sim [(1-i)/\pi] x^{-1/2} e^{ix}$ for $x \to \infty$ (which produces an outward propagating wave at infinity). A physically acceptable solution then is
\begin{equation}\label{eq:SolOutside}
\begin{aligned}
\overline{E}_{z}^{\rm (out)}(\rho, \zeta)&=-V \displaystyle\frac{H_{0}^{(1)}(\rho\sqrt{k_c^2-\zeta^2})}{H_{0}^{(1)}(a\sqrt{k_c^2-\zeta^2})}\\
&=-V \displaystyle\frac{K_{0}(\rho\sqrt{\zeta^2-k_c^2})}{K_{0}(a\sqrt{\zeta^2-k_c^2})}.
\end{aligned}
\end{equation}

We now take the first two Maxwell's equations (Faraday's and Amp\`{e}re's law), which we subsequently transform in the Fourier domain to find $\overline{B}_{\phi}(\rho,\zeta)$. The resulting coupled set of equations yields
\begin{equation}\label{eq:Bphiexpr1}
\overline{B}_{\phi}(\rho,\zeta)=\frac{ik_c}{c}\frac{1}{k_c^2-\zeta^2} \frac{\partial}{\partial \rho} \overline{E}_{z}(\rho, \zeta).
\end{equation} 
The final result reads
\begin{equation}\label{eq:Bphiexpr2}
\overline{B}_{\phi}(\rho,\zeta)=\begin{cases} \displaystyle\frac{ik_c V}{c}\frac{1}{\sqrt{k_c^2-\zeta^2}} \frac{J_1(\rho\sqrt{k_c^2-\zeta^2})}{J_0(a \sqrt{k_c^2-\zeta^2})}, \quad \quad \rho \leq a, \\\displaystyle \frac{ik_c V}{c}\frac{1}{\sqrt{\zeta^2-k_c^2}} \frac{K_1(\rho\sqrt{\zeta^2-k_c^2})}{K_0(a \sqrt{\zeta^2-k_c^2})}, \quad \quad \rho > a. \end{cases}
\end{equation}
We can then proceed to evaluate the Fourier-transformed surface current on the two metal surfaces from
\begin{equation}\label{eq:Currdensity1}
\begin{aligned}
\overline{J}_{{\rm s}z}^{\rm (in)}(\zeta)&=-\frac{1}{\mu_0}\overline{B}_{\phi}^{\rm (in)}(a, \zeta)\\
&=-\frac{i V k_c}{\zeta_c} \frac{1}{\sqrt{k_c^2-\zeta^2}}\frac{J_1(a\sqrt{k_c^2-\zeta^2})}{J_0(a\sqrt{k_c^2-\zeta^2})}
\end{aligned}
\end{equation}
and
\begin{equation}\label{eq:Currdensity2}
\begin{aligned}
\overline{J}_{{\rm s}z}^{\rm (out)}(\zeta)&=\frac{1}{\mu_0}\overline{B}_{\phi}^{\rm (out)}(a, \zeta)\\
&=-\frac{i V k_c}{\zeta_c} \frac{1}{\sqrt{\zeta^2-k_c^2}}\frac{K_1(a\sqrt{\zeta^2-k_c^2})}{K_0(a\sqrt{\zeta^2-k_c^2})}.
\end{aligned}
\end{equation}
The total current on the surface is $I(z)=2\pi a [J_{{\rm s}z}^{\rm (in)}(z)+J_{{\rm s}z}^{\rm (out)}(z)]$. We note that Eqs. \ref{eq:Currdensity1} and \ref{eq:Currdensity2} are identical to the Fourier Transforms of Eqs. (8.100) and (8.99), respectively, in \cite{CollinZucker}, with $k\to k_c$ and $\zeta_0 \to \zeta_c$ (see also Table I of \cite{FGG12013}). Furthermore, in the inverse Fourier-transform integrals of \cite{CollinZucker}, the $\zeta$-path is not indented since $k_c$ is now complex.

At this point, let us invoke explicitly the connection between the effective current and the post-processing of a current distribution located on the $z$-axis of the dipole. The effective current for $0<\rho \leq a$ can be written in terms of the vector potential $A_z(\rho,z)$ generated by the current $I_{\rm ap}(z)$ as (Eq. 19 of \cite{FGG12013} with $h=\infty$)
\begin{equation}\label{eq:currentAz}
\begin{aligned}
I_{\rm eff}(\rho, z)&=-\left(\frac{2\pi \rho}{\mu_0}\right)\frac{\partial}{\partial \rho}A_z\\
&=-\frac{\rho}{2} \int_{-\infty}^{\infty} I_{\rm ap}(z^{\prime}) \frac{\partial}{\partial \rho} \left(\frac{e^{i k_c \sqrt{(z-z^{\prime})^2 + \rho^2}}}{\sqrt{(z-z^{\prime})^2 + \rho^2}}\right)dz^{\prime} \\
&\equiv -2\pi \rho \int_{-\infty}^{\infty} I_{\rm ap}(z^{\prime}) \frac{\partial K_{\rm ap}(z-z^{\prime}, \rho)}{\partial \rho}dz^{\prime}.
\end{aligned}
\end{equation}
The Fourier transform of Eq. \ref{eq:currentAz} yields
\begin{equation}\label{eq:FTofIeff}
\overline{I}_{\rm eff}(\rho, \zeta)=-2\pi \rho \overline{I}_{\rm ap}(\zeta)\frac{\partial \overline{K}_{\rm ap}(\rho, \zeta)}{\partial \rho},
\end{equation}
which, upon substitution to the Fourier transform of PE with the approximate kernel, driven by a DFG (see Eq. (2.1) of \citep{EffCurrent2011} in free space),
\begin{equation}\label{eq:FTIapp}
(k_c^2-\zeta^2) \overline{K}_{\rm ap}(a, \zeta) \overline{I}_{\rm ap}(\zeta)=\frac{i V k_c}{\zeta_c}
\end{equation}
and using 
\begin{equation}
\overline{K}_{\rm ap}(\rho, \zeta)=\frac{1}{2\pi}K_0\left(\rho \sqrt{\zeta^2-k_c^2}\right),
\end{equation}
gives
\begin{equation}\label{eq:EqFTastI}
\overline{I}_{\rm eff}(a, \zeta)=\frac{i V k_c}{\zeta_c}w(a, \zeta),
\end{equation}
where, following the notation of Table I in \citep{FGG12013},
\begin{equation*}
w(\rho, \zeta) \equiv -\frac{2\pi \rho}{\sqrt{\zeta^2-k_c^2}}\frac{K_1(\rho\sqrt{\zeta^2-k_c^2})}{K_0(a\sqrt{\zeta^2-k_c^2})}.
\end{equation*}
Finally, invoking Eq. \ref{eq:Currdensity2} yields
\begin{equation}\label{eq:finalEqTastFT}
\overline{I}_{\rm eff}(a, \zeta)=2\pi a\, \overline{J}_{{\rm s}z}^{\rm (out)}(\zeta)
\end{equation}
and, consequently,
\begin{equation}\label{eq:finalEqTast}
I_{\rm eff}(a, z)=2\pi a\, J_{{\rm s}z}^{\rm (out)}(z).
\end{equation}
Hence, {\it the effective current at $\rho=a$ is equal to the surface current density flowing on the outer surface of the tubular antenna multiplied by its cross-sectional circumference}, a property connecting to the definition $I_{\rm eff}(a, z) \equiv 2\pi a H_{\phi}^{\rm (out)}(a,z)$ (see Eq. 8 of \cite{papakanellos2007possible}). The current distribution at $\rho=a$ can be written as $I(z) = 2\pi a[J_{{\rm s}z}^{\rm in}(z,a)+J_{{\rm s}z}^{\rm out}(z,a)]$ or, equivalently, via Eq. \ref{eq:finalEqTast} holding for a tubular dipole of infinite length, as $I(z) = 2\pi a J_{{\rm s}z}^{\rm in}(z,a)+I_{\rm eff}(z,a)$. For $|z| \to \infty$, a relation corresponding to Eq. \ref{eq:edgecondition} for the finite-length antenna is trivially satisfied, with $2\pi a J_{{\rm s}z}^{\rm in}(z\to \pm\infty,a)+I_{\rm eff}(z\to \pm \infty,a)=0$. For what concerns the finite-length antenna, our numerical results indicate that the effective current does not seem to exactly vanish near the two ends $z=\pm h$; in fact, the effective current is not bound by an edge condition of the type of Eq. \ref{eq:edgecondition}. When the exact kernel is used, application of the Wiener-Hopf method yields a square-root behavior of the current near the antenna ends \cite{ShenWu}. In short, when following the path arriving at Eq. \ref{eq:finalEqTast}, we have linked a current distribution generated by a linear current source via Eq. \ref{eq:currentAz} to a boundary condition matching the solution of Maxwell's equations inside and outside the tubular antenna. We note here that for an electrically thin dipole, the current on the inner surface of the tube is extremely small except for short ranges near the open ends \cite{KingHarrison}, a fact justifying the identification of the effective current with the current density on the {\it outer} surface in Eq. \ref{eq:finalEqTast}.  

\begin{figure*}
\begin{center}
\includegraphics[width=\textwidth]{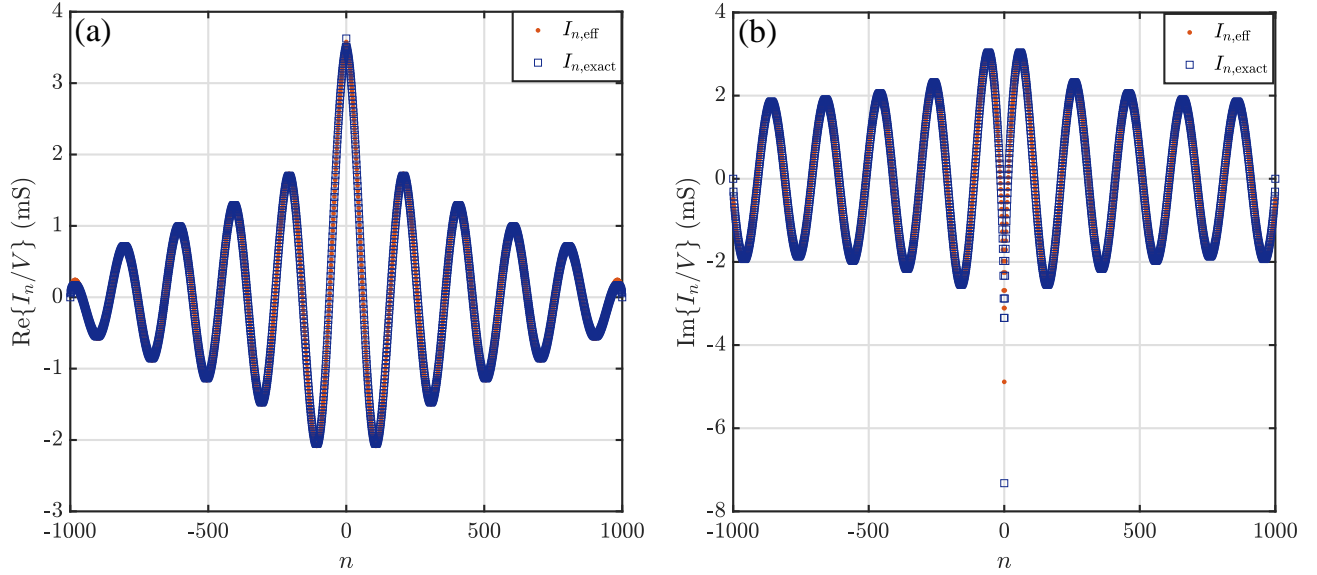}
\end{center}
\caption{{\it Alleviation of the current oscillations on a dipole larger than the wavelength (I).} Real {\bf (a)} and imaginary {\bf (b)} part of the effective current distribution $I_{n, {\rm eff}}/V$ (in orange dots), calculated from Eq. \ref{eq:IeffInFin} post-processing the oscillating current distribution $I_{n, {\rm ap}}/V$ obtained from the numerical solution of Eq. \ref{eq:Hallen} with the approximate kernel, superimposed on top of the numerical solution of Eq. \ref{eq:Hallen} with the exact kernel (in blue squares). Here, $\tan\delta=0.03$. The remaining parameters are: $h/\lambda=5$, $a/\lambda=0.02$, $N=1000$. MoM-B has been used here to generate the depicted results.}
\label{fig:largelengthsmalltanD}
\end{figure*}
\begin{figure*}
\begin{center}
\includegraphics[width=\textwidth]{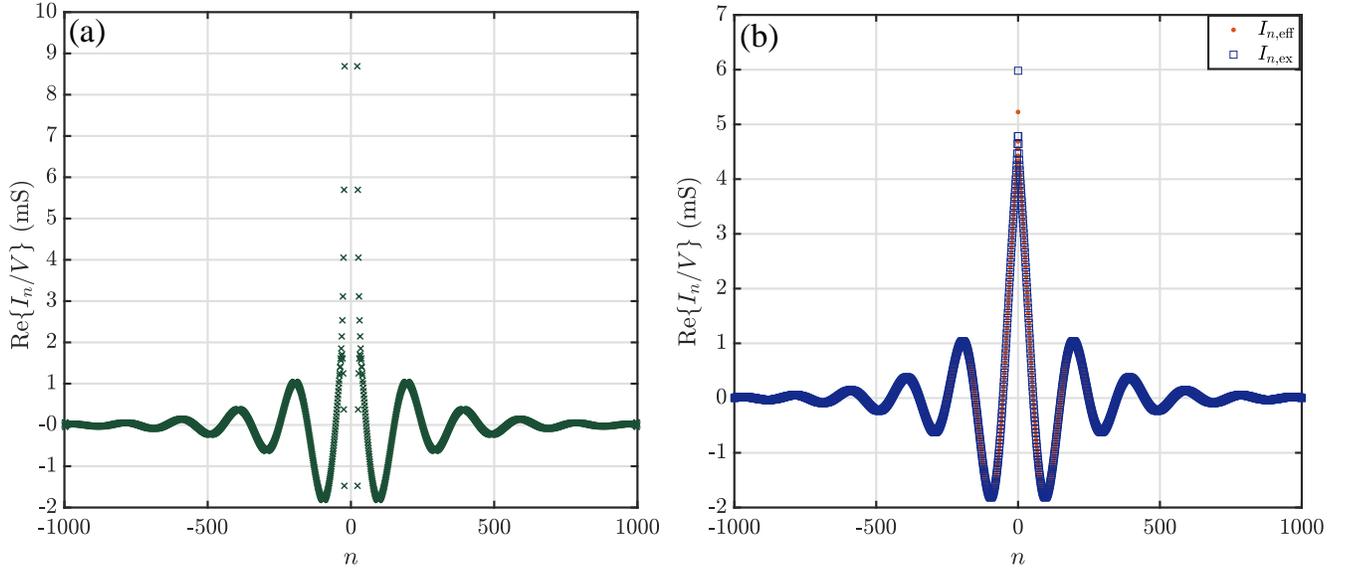}
\end{center}
\caption{{\it Alleviation of the current oscillations on a dipole larger than the wavelength (II).}  {\bf (a)} Real part of the oscillating current obtained from Eq. \ref{eq:Hallen} with the approximate kernel (omitting central oscillations for small $n$ which are out of scale), next to the corresponding effective current (in orange dots) on top of the solution $I_{n, {\rm ex}}/V$ of Eq. \ref{eq:Hallen} with the exact kernel (in blue squares) depicted in frame {\bf (b)}. At the driving point, the value of the real part of the oscillating current distribution is $I_{n=0, {\rm ap}}/V \approx 4.32\,$S [out of scale \textemdash{not} shown in (a)]. Here, $\tan\delta=0.3$. The remaining parameters are identical to those in Fig. \ref{fig:largelengthsmalltanD}. MoM-B has once more been used here to generate the depicted results.}
\label{fig:largelengthLargetanD}
\end{figure*}

\section{Numerical results for an antenna of finite length}
\label{sec:numerics}

In the present section, we will discuss the application of two different versions of the method of moments (MoM), employed to determine the expansion coefficients and the associated current for an antenna with finite length $2h$ and radius $a$. Central to both methods is the discretization in terms of $2N+1$ basis functions, together with a discretization length $z_0$ such that $N z_0 \simeq h$ for $N \gg 1$. In the first method, called {\it MoM-A} hereinafter, one seeks the unknown current as a superposition of $2N+1$ step basis functions $u_{n}(z)$ defined as \cite{Hallen2001, anastasiscnt}
\begin{equation}\label{eq:pulsesbasis}
u_n(z) =
  \begin{cases}
    1&  \text{, if } \left( n-\frac{1}{2}\right)z_0<z<\left( n+\frac{1}{2}\right)z_0\\
    0&  \text{, otherwise,}\\
  \end{cases} 
\end{equation} 
where $(2N+1)z_0=2h$,
\begin{equation}\label{eq:Ipulse}
I(z) \simeq \sum_{n=-N}^{N} I_{n} u_n(z),
\end{equation}
substitutes in Eq. \ref{eq:Hallen} and takes the inner product with the same step functions $u_{n}(z)$ to produce the system of equations
\begin{equation}\label{eq:system1}
\sum_{n=-N}^{N}A_{l-n}I_{n}^{(1)}=B_{l}^{(1)}, \quad \sum_{n=-N}^{N}A_{l-n}I_{n}^{(2)}=B_{l}^{(2)},
\end{equation}
where the coefficients $B_{l}^{(1)}$ and $B_{l}^{(2)}$ on the right-hand sides correspond to the two terms on the right-hand side of Eq. \ref{eq:Hallen} and can be derived analytically [see Eqs. (1.62a) and (1.62b) of \cite{fikiorisDylindrical} with $\beta_0 \to k_c$ and $\zeta_0 \to \zeta_c$, besides trivial notation differences]. The coefficients $A_{l}$ are elements of a Toeplitz matrix $\textrm{\bf A} \equiv [A_{kl}]$, given by the formula \cite{Hallen2001, anastasiscnt}
\begin{equation}\label{eq:Alcoeff}
A_{l}=A_{-l}=\int_{0}^{z_0}(z_0-z)[K(z+lz_0)+K(z-lz_0)]dz,
\end{equation}
where $l=0, \pm 1, \pm 2,\ldots, \pm 2N$. In Eq. \ref{eq:Alcoeff}, $K(z)$ is either the exact or the approximate kernel. The expansion coefficients $I_{n}$ are then given by the superposition $I_{n}=I_{n}^{(1)}+C I_{n}^{(2)}$, with $I_{\pm N}=0$, after having determined the constant $C$ by the edge condition given in Eq. \ref{eq:edgecondition}. MoM-A is the principal method employed in \cite{Hallen2001} to ascertain the existence of unphysical oscillations. 

In the other numerical variant, which is a {\it collocation} method we term {\it MoM-B}, we are  seeking the current distribution in the form 
\begin{equation}\label{eq:Itriang}
I(z) \simeq \sum_{n=-(N-1)}^{N-1} I_{n} t(z-nz_0),
\end{equation}
where now $N z_0=h$. Here, $t(z)$ is the triangular basis function of Eq. \ref{eq:triangular}. Demanding that Eq. \ref{eq:Hallen} holds at the points $z=l z_0$, with $l=0, \pm 1,\ldots,\pm N$, produces a similar system of equations to that of Eq. \ref{eq:system1} for $I_{n}=I_{n}^{(1)}+C I_{n}^{(2)}$ \cite{Hallen2001}, in which the coefficients $A_{l}$ are also given by Eq. \ref{eq:Alcoeff}, whereas the coefficients on the right-hand side are now $B_{l}^{(1)}=[iV z_0/(2\zeta_c)]\sin(k_c|l|z_0)$ and $B_{l}^{(2)}=z_{0} C\cos(k_c l z_0)$. The constant $C$ is once more determined by the edge condition of Eq. \ref{eq:edgecondition}.

These two methods, MoM-A and MoM-B, are used for the numerical investigation of the finite-length antenna in a conducting medium, producing the expansion coefficients $I_{n}$ subject to the choice of a particular kernel (either exact or approximate). Similarly to Eq. \ref{eq:IeffIn}, we define the effective current distribution at $z=m z_0$ as 
\begin{equation}\label{eq:IeffInFin}
\begin{aligned}
&I_{\rm eff}(\rho, mz_0, z_0)=\frac{1}{2i \sin(k_c z_0)}\\
&\times \sum_{n=-(N-1)}^{N-1}[f_{n+1}-2\cos(k_c z_0) f_{n}+f_{n-1}]I_{n, {\rm ap}},
\end{aligned}
\end{equation}
where $I_{n, {\rm ap}}$ originate from the rapidly-oscillating current distribution solving Eq. \ref{eq:Hallen} with the approximate kernel (for $N \gg h/a$) and $f_{n}\equiv\exp[ik_c\sqrt{(nz_0-mz_0)^2+\rho^2}]$ \footnote{Strictly speaking, Eq. \ref{eq:IeffInFin} is different from $2\pi \rho$ times the exact magnetic field due to the current distribution of Eq. \ref{eq:Itriang}. The functions $s(z-nz_0)$ and $t(z-nz_0)$, however, are almost identical for small $z_0$.}. For the case of pulse basis functions $u_n(z)$ with very small $z_0$ used in MoM-A, we instead approximate the magnetic field by that of a Hertzian dipole of finite length to produce an approximate formula for $I_{\rm eff}(\rho,z)$. Eq. \ref{eq:IeffInFin} has also been employed to alleviate oscillations arising in the solution of Eq. \ref{eq:Hallen} with the approximate kernel for a finite-gap generator \cite{FGG12013}.
\begin{figure}
\begin{center}
\includegraphics[width=0.5\textwidth]{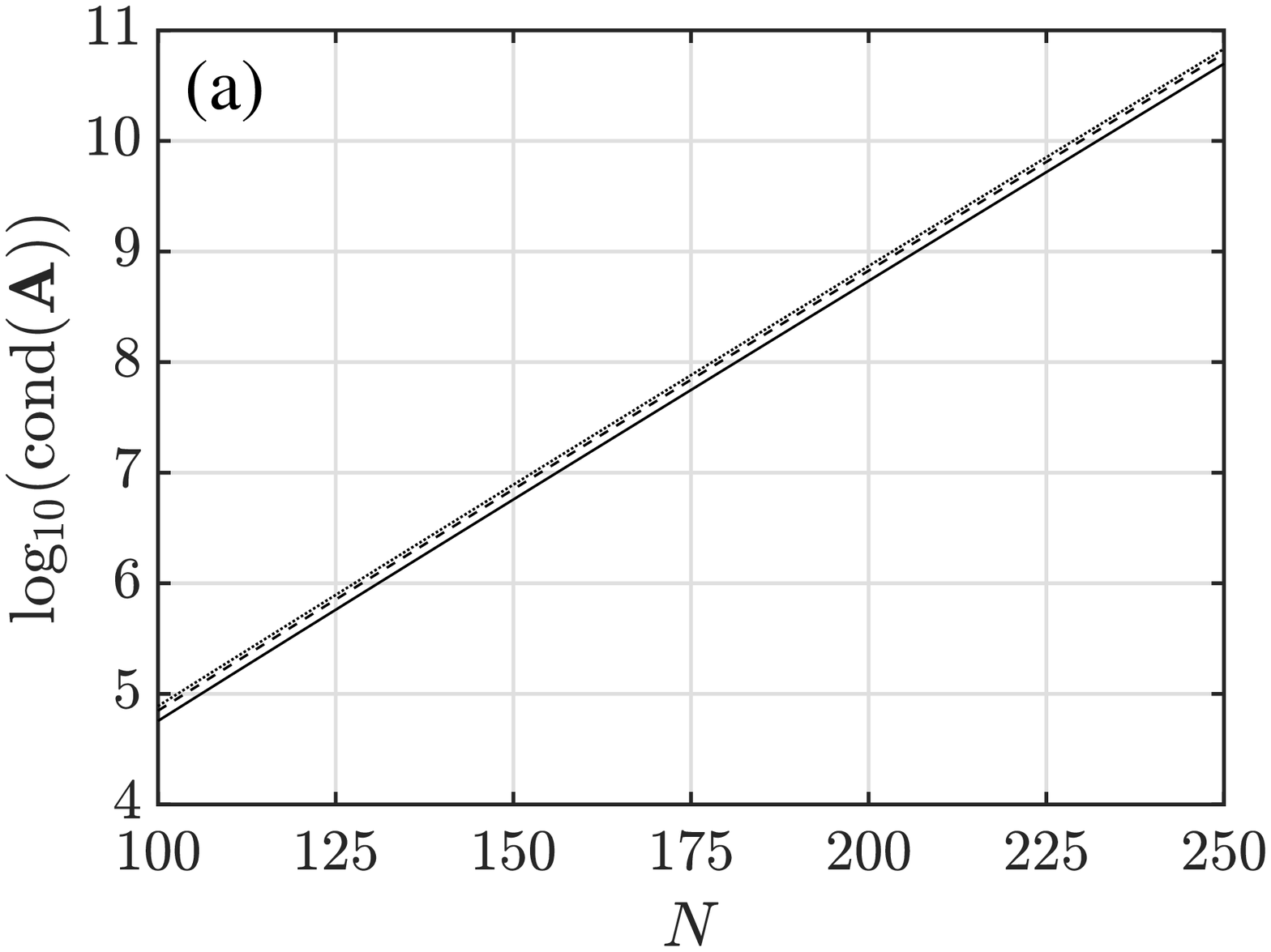}
\includegraphics[width=0.5\textwidth]{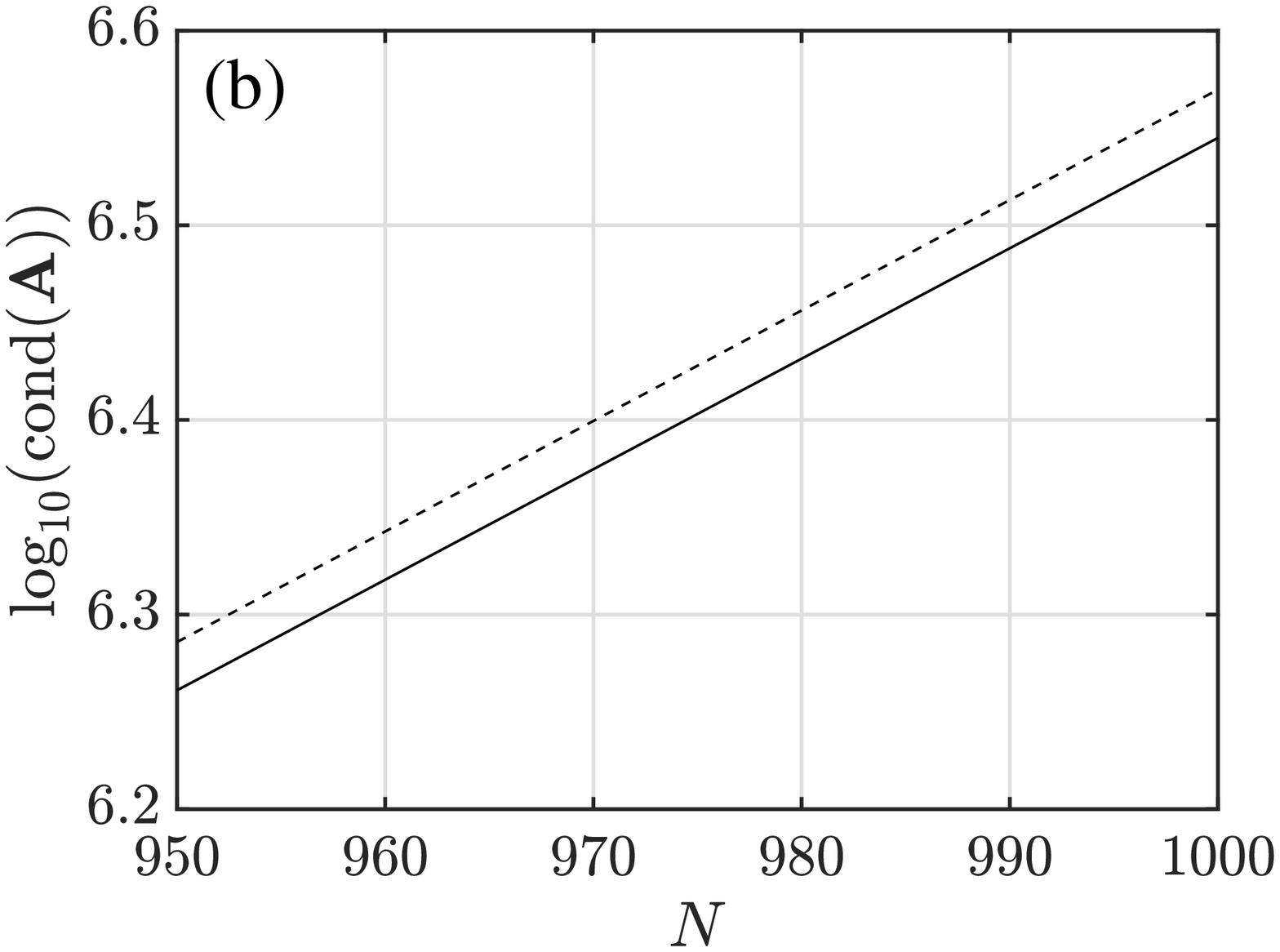}
\end{center}
\caption{{\it Ill-conditioning of matrix $[A_{kl}]$ with the approximate kernel for varying medium conductivity.} Logarithm of the $2-$norm condition number as a function of the matrix size $N$. In {\bf (a)}, $h/\lambda=0.25$, $a/\lambda=0.007022$, and $\tan \delta=0, 7.19, 35.95$ as depicted by the dotted, dashed and solid lines respectively. In {\bf (b)}, $h/\lambda=5$, $a/\lambda=0.02$ and $\tan\delta=0.03, 0.07$ as depicted by the dashed and solid lines, respectively.}
\label{fig:condnum}
\end{figure}

Fig. \ref{fig:approx150} evidences a dramatic change in the behavior of ${\rm Re}\{I_{n, {\rm ap}}/V\}$ once a finite conductivity is introduced [compare frames (a) and (c)], with the central oscillations, inherited from the imaginary part, taking over the side oscillations in the mS range. The imaginary parts of the oscillatory currents, on the other hand, remain virtually identical. For the infinite antenna, the effective current $I_{\rm eff}(0, nz_0, z_0)$ asymptotically equals the current $I_n$ obtained by the numerical method we have described in Sec. \ref{sec:effectiveinf} \cite{EffCurrent2011, EffCurrent2013}. For the antenna of finite length, this property is verified by comparing the results from the application of MoM-A and the asymptotic formula \ref{eq:asymptoticIeff}. The only difference between Figs. \ref{fig:approx150}(c) and \ref{fig:effectivetwodist}(a) amounts to an overall factor of $2$; this discrepancy can be also traced in the asymptotic analysis of the infinite antenna using MoM-A and MoM-B (see Sec. VIII of \cite{Hallen2001}). Moving away from the axis, the oscillations wane gradually to produce a current distribution which compares well with the solution of Eq. \ref{eq:Hallen} with the exact kernel, as we can see in Fig. \ref{fig:effectivetwodist}(b). 

In Fig. \ref{fig:largelengthsmalltanD}, we depict the current distribution on a dipole with $h=5\lambda$, embedded in a surrounding medium with small conductivity ($\tan\delta=0.03$). The oscillating expansion coefficients and the associated line current from the approximate-kernel formulation are once more smoothed and in very good agreement with the solution of HE with the exact kernel, as we can see in frames (a) and (b) for the real and imaginary parts, respectively. With a growing loss tangent, more intense central oscillations are inherited to the real part of $I_{n, {\rm ap}}$ from the imaginary part, as we can observe in Fig. \ref{fig:largelengthLargetanD}(a). The intense oscillations near the driving point, with a difference of several orders of magnitude from the rest of the current distribution at larger $|z|$, are once again remedied in Fig. \ref{fig:largelengthLargetanD}(b). The remedy comes along with a decaying current distribution which gradually loses its side oscillations in favor of a more pronounced central peak. 

Let us now turn to the numerical stability of the two methods employed in our analysis. We stress once more that the displayed unphysical oscillations are not due to matrix ill-conditioning, which is nevertheless an important effect typically ocurring in Fredholm equations of the first kind with an analytic kernel \cite{AtkinsonKEbook, Groetsch2007, Kress}. Since the effective current is expected to be small while the oscillating current coefficients $I_{n, {\rm ap}}$ \textemdash{arising} from a highly ill-conditioned system \textemdash{are} typically orders of magnitude larger, our final results are produced by subtractive cancellation (see the concluding remarks of \cite{GMTFikiorisTsitsas}). This can be immediately traced back to the explicit form of Eq. \ref{eq:IeffIn} relating the effective current to rapidly oscillating quantities of very large magnitude. The aforementioned cancellation is tantamount to a great sensitivity of the final results to roundoff errors in the intermediate quantities $I_{n,{\rm ap}}$; these quantities, in turn, deteriorate rapidly because of the growing condition number of the matrix $[A_{ln}]$ in the system of equations \ref{eq:system1}.

Increasing the dimension $N$ of the matrix $\textrm{\bf A}$ with elements $A_{ln} \equiv A_{|l-n|}$, as defined in Eq. \ref{eq:Alcoeff}, entails a significant growth of both $I_{n, {\rm ap}}$ and the matrix condition number. The latter scales {\it exponentially} with $N$ (see \cite{Mittra1975}), as we observe in Fig. \ref{fig:condnum}, where we depict the $2-$norm condition number for various values of $N$, $h/\lambda$, $a/\lambda$ and $\tan\delta$, corresponding to the short and long dipoles of Figs. \ref{fig:approx150} and \ref{fig:largelengthsmalltanD} (see also Fig. 3 and Sec. VII of \cite{FrillFikLionas}). The two graphs evidence an exponential growth of the matrix $A_{kl}$ condition number with $N$ to an excellent degree of approximation for any value of the ambient conductivity. We also find that the important parameter in determining the matrix ill-conditioning is the ratio $h/a$ (with unphysical oscillations arising for $N\gg h/a$), while, for a fixed value of $h/a$, the condition number drops with increasing conductivity of the surrounding medium.

Further to Eq. (4.13) of \citep{EffCurrent2011}, for $\rho=a$ and $z_0 \to 0$, the effective current for the infinite antenna, surrounded by a medium of a finite conductivity, exhibits a singular behavior, according to
\begin{equation}\label{eq:logsing}
\frac{I_{\rm eff}^{(\infty)}(a,z,z_0 \to 0)}{V} \sim -i \frac{2 k_c a}{\zeta_c} \ln \frac{1}{|z|},\,\, z \to 0,
\end{equation}  
reproducing the small-$z$ logarithmic singularity of the exact current $I_{\rm ex}^{(\infty)}(a,z, z_0 \to 0)$ apart from an overall factor of two. The appearance of this factor has a notable physical significance: near the driving point, the charge densities on the outer and inner surfaces diverge in a similar fashion, rendering the effective current density (which, we have shown, equals density on the {\it outer} surface only) a half of the corresponding exact current (defined as the sum of {\it both} densities). As we move away from the gap, however, the inner distribution rapidly decays, while the two current distributions approach each other (see Sec. IIIC of \cite{TastsoglouII}). Extrapolating now to the antenna of finite length $2h$ and finite $z_0$, the asymptotic formula given in Eq. \ref{eq:logsing} explains the pronounced peak of the current distribution at $n=0$ for a growing $\tan\delta$, similar to the one we observe in Figs. \ref{fig:effectivetwodist}(b) and Fig. \ref{fig:largelengthLargetanD}(b). This is in agreement with the experimental results for a dipole in a homogeneous conducting medium (where $\epsilon \approx 75 \epsilon_0$) depicted in Fig. 2 of \cite{PopovicConducting} and Fig. 11 of \cite{IizukaKing} (see also Ch. 6 of \cite{PopovicBook}). 

\section{Concluding discussion}
\label{sec:concldisc}

In conclusion, we have discussed a remedy for the unphysical oscillations arising in the current distributions of linear antennas with an analytic kernel which are not due to roundoff errors or to matrix ill-conditioning effects. Rather, they are a consequence of the nonsolvability of the underlying equation, independent of the specifics of the numerical method employed: oscillations occur with different basis and testing functions. We emphasize once more that the approximate-kernel formulation amounts to seeking a line current on the $z$-axis ($\rho=0$) rather than the actual current provided by the exact-kernel formulation on the antenna surface ($\rho=a$). The effective current aims to join the two pictures, producing a smooth current distribution at $\rho=a$, in very good agreement with the actual current obtained from the exact kernel, and a rapidly oscillating waveform on the $z$-axis, reproducing the unphysical oscillations due to the nonsolvability of HE with the approximate kernel. 

What insight does now the extension to a medium of finite conductivity offer? As we have seen from the asymptotic analysis and the numerics, intense central oscillations are inherited from the imaginary to the real part when the approximate kernel is used as a result of a nonzero loss tangent. The effective current accommodates such a change on the nature of the displayed oscillations producing a smooth current distribution for both the real and imaginary parts at $\rho=a$ (the antenna surface). This is not a mere numerical coincidence; it is due to the equality between a solution of Maxwell's equations and their associated boundary conditions, and the field generated by a linear current on the antenna axis due to the approximate-kernel formulation as a result of a {\it post-processing} method. The solution of HE with the exact kernel, which is based on the actual total current, comes to corroborate this connection, comparing very well with the effective current on the surface of the antenna {\it away from the driving point}.

A final comment regarding the applicability of Maxwell's equations is in order. The small-$z$ logarithmic singularity of the effective current for $z_0 \to 0$, that we met in Sec. \ref{sec:numerics}, only pertains to a mathematical limit; as such, it is not constrained by the requirement of a physically acceptable field. For any finite $z_0$, however, the effective-current distribution represents a {\it true field} produced by individual sources of length $2z_0$ (sinusoidal currents along the $z$-axis). This field satisfies Maxwell's equations and has no singularity whatsoever in $0<\rho \leq a$ (see Sec. 4 of \cite{EffCurrent2011}). The effective-current formulation resembles MAS with regard to the appearance of oscillations in the fictitious source. Hence, the asymptotic analysis in Sec. \ref{sec:effectiveinf} and the extension to a medium of finite conductivity is relevant to the study of near fields generated by rapidly oscillating currents (which have been proven explicitly distinct from roundoff effects) in scattering problems \cite{Psarros2007, Andrianesis2012}.   

\bibliography{BibliographyREV}

\end{document}